# DECAYS OF BOTTOM MESONS EMITTING TENSOR MESON IN FINAL STATE USING ISGW II MODEL

Neelesh Sharma, Rohit Dhir* and R.C. Verma
Department of Physics, Punjabi University,
Patiala-147 002, INDIA;
*School of Physics and Material Science,
Thapar University, Patiala-147 004, INDIA.

**Abstract**

In this paper, we investigate phenomenologically two-body weak decays of the bottom mesons emitting pseudoscalar/vector meson and a tensor meson. Form factors are obtained using the improved ISGW II model. Consequently, branching ratios for the CKM-favored and CKM-suppressed decays are calculated.

PACS no (s): 13.25.Hw, 12.39.Jh, 12.39.St

# I. INTRODUCTION

Experimental results are available for the branching ratios of several *B*-meson decay modes. Many theoretical works have been done to understand exclusive hadronic *B* decays in the framework of the generalized factorization, QCD factorization or flavor SU(3) symmetry. Weak hadronic decays of the *B*-mesons are expected to provide a rich phenomenology yielding a wealth of information for testing the standard model and for probing strong interaction dynamics. However, these decays involve nonperturbative strong processes which cannot be calculated from the first principles. Thus, phenomenological approaches [1-5] have generally been applied to study them using factorization hypothesis. It involves expansion of the transition amplitudes in terms of a few invariant form factors which provide essential information on the structure of the mesons and the interplay of the strong and weak interactions. This scheme has earlier been employed to study the weak hadronic decays of *B*-meson to *s*-wave mesons [5-12]. *B*-mesons, being heavy, can also emit heavier mesons such as *p*-wave mesons, which have attracted theoretical attention recently. However, there exist only a few works on the hadronic *B* decays [13-17] that involve a tensor meson in the final state using the frameworks of flavor SU(3) symmetry and the generalized factorization. In the next few years new experimental data on rare decays of *B* mesons would become available from the *B* factories such as KEK, Belle, Babar, BTeV, LHC. It is expected that improved measurements or new bounds will be obtained on the branching ratios for various decay modes and many decay modes with small branching ratios may also be observed for the first time.

In this paper, we analyze two-body hadronic decays of $B^-$, $\bar{B}^0$ and $\bar{B}_s^0$ mesons to pseudoscalar ($P$ ($0^-$)) /vector ($V$ ($1^-$)) and tensor ($T$ ($2^+$)) mesons in the final state, for whom the experiments have provided the following branching ratios [18,19]:

$$
\begin{aligned}
B(B^- \to \pi^- D_2^0) &= (7.8 \pm 1.4) \times 10^{-4}, \\
B(B^- \to \pi^- f_2) &= (8.2 \pm 2.5) \times 10^{-6}, \\
B(B^- \to K^- f_2) &= (1.3^{+0.4}_{-0.5}) \times 10^{-6}, \\
B(B^- \to \eta K_2^-) &= (9.1 \pm 3.0) \times 10^{-6}, \\
B(\bar{B}^0 \to \eta \bar{K}_2^0) &= (9.6 \pm 2.1) \times 10^{-6}, \\
B(\bar{B}^0 \to \bar{D}^0 f_2) &= (1.2 \pm 0.4) \times 10^{-4}, \\
B(\bar{B}^0 \to \phi \bar{K}_2^0) &= (7.8 \pm 1.3) \times 10^{-6}, \\
B(\bar{B}^0 \to \pi^{\mp} a_2^{\pm}) &< 3.0 \times 10^{-4}, \\
B(B^- \to \pi^- \bar{K}_2^0) &< 6.9 \times 10^{-6}, \\
B(\bar{B}^0 \to D_s^- a_2^+) &< 1.9 \times 10^{-4}, \\
B(\bar{B}^0 \to \pi^+ K_2^-) &< 1.8 \times 10^{-5}, \\
B(\bar{B}^0 \to \pi^- D_2^+) &< 2.2 \times 10^{-3}, \\
B(B^- \to \rho^- D_2^0) &< 4.7 \times 10^{-3},
\end{aligned} \quad (1)
$$

$$B(B^- \to \rho^- K_2^-) < 1.5 \times 10^{-3},$$
$$B(B^- \to \phi K_2^-) < 3.4 \times 10^{-3},$$
$$B(B^- \to \rho^0 a_2^-) < 7.2 \times 10^{-3},$$
$$B(\bar{B}^0 \to D_s^+ a_2^-) < 2.0 \times 10^{-4},$$
$$B(\bar{B}^0 \to \rho^- D_2^+) < 4.9 \times 10^{-3},$$
$$B(\bar{B}^0 \to \rho^0 \bar{K}_2^0) < 1.1 \times 10^{-3}.$$

In general, *W*-annihilation and *W*-exchange diagrams may also contribute to these decays under consideration. Normally, such contributions are expected to be suppressed due to the helicity and color arguments and are neglected in this work.

The paper is organized as follows: In Sec. II, we present meson spectroscopy. Methodology for calculating $B \to PT$ and $B \to VT$ is provided in Sec. III. Sec. IV deals with numerical results and discussions. Summary and conclusions are given in the last section.

## II. MESON SPECTROSCOPY

Experimentally [18], the tensor meson sixteen-plet comprises of an isovector $a_2(1.318)$, strange isospinor $K_2^*(1.429)$, charm SU(3) triplet $D_2^*(2.457)$, $D_{s2}^*(2.573)$ and three isoscalars $f_2(1.275)$, $f_2'(1.525)$ and $\chi_{c2}(3.555)$. These states behave well with respect to the quark model assignments, though the spin and parity of the charm isosinglet $D_{s2}^*(2.573)$ remain to be confirmed. The numbers given within parentheses indicate mass (in GeV units) of the respective mesons. $\chi_{c2}(3.555)$ is assumed to be pure $(c\bar{c})$ state, and mixing of the isoscalar states is defined as:

$$f_2(1.275) = \frac{1}{\sqrt{2}}(u\bar{u} + d\bar{d})\cos\phi_T + (s\bar{s})\sin\phi_T,$$
$$f_2'(1.525) \frac{1}{\sqrt{2}}(u\bar{u} + d\bar{d})\sin\phi_T - (s\bar{s})\cos\phi_T, \qquad (2)$$

where $\phi_T = \theta(ideal) - \theta_T(physical)$ and $\theta_T(physical) = 27°$ [18].

Similarly, for $\eta$ and $\eta'$ states of well established pseudoscalar sixteen-plet, we use

$$\eta(0.547) = \frac{1}{\sqrt{2}}(u\bar{u} + d\bar{d})\sin\phi_P - (s\bar{s})\cos\phi_P,$$

$$\eta'(0.958) = \frac{1}{\sqrt{2}}(u\bar{u} + d\bar{d})\cos\phi_P + (s\bar{s})\sin\phi_P, \qquad (3)$$

where $\phi_p = \theta(ideal) - \theta_p(physical)$ and we take $\theta_P(physical) = -15.4°$ [18]. $\eta_c$ is taken as

$$\eta_c(2.979) = (c\bar{c}). \qquad (4)$$

Similarly, for $\omega$ and $\phi$ states of well established pseudoscalar sixteen-plet, we use

$$\omega(0.783) = \frac{1}{\sqrt{2}}(u\bar{u} + d\bar{d})\cos\phi_V + (s\bar{s})\sin\phi_V,$$

$$\phi(1.019) = \frac{1}{\sqrt{2}}(u\bar{u} + d\bar{d})\sin\phi_V - (s\bar{s})\cos\phi_V, \qquad (5)$$

where $\phi_V = \theta(ideal) - \theta_V(physical)$ and we take $\theta_V(physical) = 39°$ [18]. $J/\psi$ is taken as

$$J/\psi(3.097) = (c\bar{c}). \qquad (6)$$

### III. METHODOLOGY

#### A. Weak Hamiltonian

For bottom changing $\Delta b = 1$ decays, the weak Hamiltonian involves the bottom changing current,

$$J_\mu = (\bar{c}b)V_{cb} + (\bar{u}b)V_{ub}, \qquad (7)$$

where $(\bar{q}_i q_j) \equiv \bar{q}_i \gamma_\mu (1-\gamma_5) q_j$ denotes the weak *V-A* current. QCD modified weak Hamiltonian is then given below:

i)   for decays involving $b \to c$ transition,

$$H_W = \frac{G_F}{\sqrt{2}} \{V_{cb}V_{ud}^*[a_1(\bar{c}b)(\bar{d}u) + a_2(\bar{d}b)(\bar{c}u)] + $$
$$V_{cb}V_{cs}^*[a_1(\bar{c}b)(\bar{s}c) + a_2(\bar{s}b)(\bar{c}c)] + \quad (8a)$$
$$V_{cb}V_{us}^*[a_1(\bar{c}b)(\bar{s}u) + a_2(\bar{s}b)(\bar{c}u)] + $$
$$V_{cb}V_{cd}^*[a_1(\bar{c}b)(\bar{d}c) + a_2(\bar{d}b)(\bar{c}c)]\},$$

ii)  for decays involving $b \to u$ transition,

$$H_W = \frac{G_F}{\sqrt{2}} \{V_{ub}V_{cs}^*[a_1(\bar{u}b)(\bar{s}c) + a_2(\bar{s}b)(\bar{u}c)] + $$
$$V_{ub}V_{ud}^*[a_1(\bar{u}b)(\bar{d}u) + a_2(\bar{d}b)(\bar{u}u)] + \quad (8b)$$
$$V_{ub}V_{us}^*[a_1(\bar{u}b)(\bar{s}u) + a_2(\bar{s}b)(\bar{u}u)] + $$
$$V_{ub}V_{cd}^*[a_1(\bar{u}b)(\bar{d}c) + a_2(\bar{d}b)(\bar{u}c)]\},$$

where $\bar{q}_i q_j \equiv \bar{q}_i \gamma_\mu (1-\gamma_5) q_j$ denotes the weak V-A current and $V_{ij}$ are the well-known CKM matrix elements, $a_1$ and $a_2$ are the QCD coefficients. By factorizing matrix elements of the four-quark operator contained in the effective Hamiltonian (6), one can distinguish three classes of decays [20]:

- class I transition <u>caused by color favored diagram</u>: the corresponding decay amplitudes are proportional to $a_1$, where $a_1(\mu) = c_1(\mu) + \frac{1}{N_c} c_2(\mu)$, and $N_c$ is the number of colors.
- class II transition <u>caused by color suppressed diagram</u>: the corresponding decay amplitudes in this class are proportional to $a_2$ i.e. for the color suppressed modes $a_2(\mu) = c_2(\mu) + \frac{1}{N_c} c_1(\mu)$.
- class III transition <u>caused by both color favored and color suppressed diagrams</u>: these decays experience the interference of color favored and color suppressed diagrams.

We follow the general convention of large $N_c$ limit to fix the QCD coefficients $a_1 \approx c_1$ and $a_2 \approx c_2$, where [20,21]:

$$c_1(\mu) = 1.12 \ , \ c_2(\mu) = -0.26 \text{ at } \mu \approx m_b^2. \quad (9)$$

## B. Decay Amplitudes and Rates

### a) $B \to PT$ Decay:

The decay rate formula for $B \to PT$ decays is given by

$$\Gamma(B \to PT) = \left(\frac{m_B}{m_T}\right)^2 \frac{p_C^5}{12\pi m_T^2} |A(B \to PT)|^2, \quad (10)$$

where $p_C$ is the magnitude of the three-momentum of the final-state particle in the rest frame of $B$-meson and $m_B$ and $m_T$ denote masses of the $B$-meson and tensor meson, respectively.

The factorization scheme in general expresses the weak decay amplitude as the product of matrix elements of weak currents (up to the weak scale factor of $\frac{G_F}{\sqrt{2}} \times CKM$ elements $\times$ QCD factor),

$$\langle PT | H_W | B \rangle \approx \langle P | J^\mu | 0 \rangle \langle T | J_\mu | B \rangle + \langle T | J^\mu | 0 \rangle \langle P | J_\mu | B \rangle. \quad (11)$$

However, the matrix elements $\langle T(q_\mu) | J_\mu | 0 \rangle$ vanish due to the tracelessness of the polarization tensor $\epsilon_{\mu\nu}$ of spin 2 meson and the auxiliary condition $q^\mu \epsilon_{\mu\nu} = 0$ [19]. Remaining matrix elements are expressed as:

$$\langle P(k_\mu) | J_\mu | 0 \rangle = -i f_P k_\mu,$$
$$\langle T(P_T) | J_\mu | B(P_B) \rangle = i h \epsilon_{\mu\nu\lambda\rho} \epsilon^{*\nu\alpha} P_{B\alpha} (P_B + P_T)^\lambda (P_B - P_T)^\rho + k \epsilon^*_{\mu\nu} P_B^\nu$$
$$+ b_+ (\epsilon^*_{\alpha\beta} P_B^\alpha P_B^\beta)[(P_B + P_T)_\mu + b_- (P_B - P_T)_\mu], \quad (12)$$

in the ISGW model [3] which yields

$$\langle PT | H_W | B \rangle = -i f_P (\epsilon^*_{\mu\nu} P_B^\mu P_B^\nu) F^{B \to T}(m_P^2), \quad (13)$$

where

$$F^{B \to T}(m_P^2) = k(m_P^2) + (m_B^2 - m_T^2) b_+ (m_P^2) + m_P^2 b_- (m_P^2). \quad (14)$$

Thus

$$A(B \to PT) = \frac{G_F}{\sqrt{2}} \times (CKM\ factors \times QCD\ factors \times CG\ factors) \times f_P F^{B \to T}(m_P^2). \quad (15)$$

## b) $B \to VT$ Decay:

The decay rate formula for $B \to VT$ is,

$$\Gamma(B \to VT) = \frac{G_F^2}{48\pi m_T^4} f_V^2 [\alpha |\vec{p}_V|^7 + \beta |\vec{p}_V|^5 + \gamma |\vec{p}_V|^3], \qquad (16)$$

where $|\vec{p}_V|$ is the magnitude of the three-momentum of the final-state particle $V$ or $T$ ($|\vec{p}_V| = |\vec{p}_T|$) in the rest frame of $B_c$ meson. $\alpha$, $\beta$ and $\gamma$, respectively, are quadratic functions of the form factors, are given by

$$\begin{aligned} \alpha &= 8 m_{B_c}^4 b_+^2, \\ \beta &= 2 m_{B_c}^2 [6 m_V^2 m_T^2 h^2 + 2(m_{B_c}^2 - m_T^2 - m_V^2) k b_+ + k^2], \\ \gamma &= 5 m_T^2 m_V^2 k^2. \end{aligned} \qquad (17)$$

Here also the decay amplitude can be expressed as the product of matrix elements of weak currents (up to the weak scale factor of $\frac{G_F}{\sqrt{2}} \times$ CKM elements×QCD factor):

$$\langle VT | H_W | B \rangle \sim \langle V | J^\mu | 0 \rangle \langle T | J_\mu | B \rangle, \qquad (18)$$

due to vanishing $\langle T | J_\mu | 0 \rangle$ matrix element. Here

$$\langle V | J_\mu | 0 \rangle = \epsilon_\mu^* m_V f_V, \qquad (19)$$

where $\epsilon_\mu^*$ and $f_V$ denote the polarization four-vector of $V$ and the decay constant of the vector meson. Relations (14) and (21) yields

$$\langle VT | H_W | B_c \rangle = m_V f_V F^{B_c \to T}(m_V^2), \qquad (20)$$

where

$$F_{\alpha\beta}^{B_c \to T} = \epsilon_\mu^* (P_{B_c} + P_T)_\rho [ih\varepsilon^{\mu\nu\rho\sigma} g_{\alpha\nu} (P_V)_\beta (P_V)_\sigma + k \delta_\alpha^\mu \delta_\beta^\rho + b_+ (P_V)_\alpha (P_V)_\beta g^{\mu\rho}], \qquad (21)$$

leading to

$$A(B \to VT) = \frac{G_F}{\sqrt{2}} \times (CKM\ factors \times QCD\ factors) \times m_V f_V \epsilon^{*\alpha\beta} F_{\alpha\beta}^{B_c \to T}(m_V^2). \qquad (22)$$

## C. Form Factors in the ISGW II Model

The form factors have the following expressions in the ISGW II quark model, for $B \to T$ transitions [3]:

$$k = \frac{m_d}{\sqrt{2}\beta_B}(1+\tilde{\omega}) \, F_5^{(k)},$$

$$b_+ + b_- = \frac{m_d}{4\sqrt{2}m_d m_b \tilde{m}_B \beta_B} \frac{\beta_T^2}{\beta_{BT}^2}\left(1 - \frac{m_d}{2\tilde{m}_B}\frac{\beta_T^2}{\beta_{BT}^2}\right) F_5^{(b_+ + b_-)},$$

(23)

$$b_+ - b_- = -\frac{m_d}{\sqrt{2}m_b \tilde{m}_T \beta_B}\left(1 - \frac{m_d m_b}{2\mu_+ \tilde{m}_B}\frac{\beta_T^2}{\beta_{BT}^2} + \frac{\beta_T^2}{4\beta_{BT}^2}\left(1 - \frac{m_d}{2\tilde{m}_B}\frac{\beta_T^2}{\beta_{BT}^2}\right)\right) F_5^{(b_+ - b_-)},$$

where

$$F_5^{(k)} = F_5 \left(\frac{\bar{m}_B}{\tilde{m}_B}\right)^{-1/2}\left(\frac{\bar{m}_T}{\tilde{m}_T}\right)^{1/2},$$

$$F_5^{(b_+ + b_-)} = F_5 \left(\frac{\bar{m}_B}{\tilde{m}_B}\right)^{-5/2}\left(\frac{\bar{m}_T}{\tilde{m}_T}\right)^{1/2},$$

(24)

$$F_5^{(b_+ - b_-)} = F_5 \left(\frac{\bar{m}_B}{\tilde{m}_B}\right)^{-3/2}\left(\frac{\bar{m}_T}{\tilde{m}_T}\right)^{-1/2},$$

The $t(\equiv q^2)$ dependence is given by

$$\tilde{\omega} - 1 = \frac{t_m - t}{2\bar{m}_B \bar{m}_T},$$

(25)

and the common scale factor

$$F_5 = \left(\frac{\tilde{m}_T}{\tilde{m}_B}\right)^{1/2}\left(\frac{\beta_T \beta_B}{\beta_{BT}^2}\right)^{5/2}\left[1 + \frac{1}{18}h^2(t_m - t)\right]^{-3},$$

(26)

where

$$h^2 = \frac{3}{4m_b m_q} + \frac{3m_d^2}{2\bar{m}_B \bar{m}_T \beta_{BT}^2} + \frac{1}{\bar{m}_B \bar{m}_T}\left(\frac{16}{33 - 2n_f}\right)\ln\left[\frac{\alpha_S(\mu_{QM})}{\alpha_S(m_q)}\right],$$

(27)

and

$$\beta_{BT}^2 = \frac{1}{2}\left(\beta_B^2 + \beta_T^2\right).$$

$\tilde{m}$ is the sum of the mesons constituent quarks masses, $\bar{m}$ is the hyperfine averaged physical masses, $n_f$ is the number of active flavors, which is taken to be five in the present case, $t_m = (m_B - m_T)^2$ is the maximum momentum transfer and

$$\mu_+ = \left(\frac{1}{m_d} + \frac{1}{m_b}\right)^{-1}, \qquad (28)$$

Here, $m_d$ is the spectator quark mass in the decaying particle. For $B_s \to T$ transitions, $m_d$ is replaced with $m_s$. We take the following constituent quark masses (in GeV):

$$m_u = m_d = 0.33,\ m_s = 0.55,\ m_c = 1.82,\ m_b = 5.20, \qquad (29)$$

which are taken from the ISGW II model [3] which treats mesons as composed of the constituent quarks. Values of the parameter $\beta$ for different $s$-wave and $p$-wave mesons are given in the Table I. We obtain the form factors describing $B \to T$ transitions which are given in Table II at $q^2 = t_m$.

## IV. NUMERICAL RESULTS AND DISCUSSIONS

For numerical calculations, we use the following values of the decay constants (given in GeV) of the pseudoscalar [13, 18, 21, 22] and vector mesons:

$$f_\pi = 0.131,\ f_K = 0.160,\ f_D = 0.223,\ f_{D_s} = 0.294,$$
$$f_\eta = 0.133,\ f_{\eta'} = 0.126 \text{ and } f_{\eta_c} = 0.400. \qquad (30)$$

and

$$f_\rho = 0.221,\ f_{K^*} = 0.220,\ f_{D^*} = 0.245,\ f_{D_s^*} = 0.273,$$
$$f_\omega = 0.195,\ f_\phi = 0.229,\ f_{J/\psi} = 0.411. \qquad (31)$$

We calculate branching ratios of $B$-meson decays in CKM-favored and CKM-suppressed modes involving $b \to c$ and $b \to u$ transitions. The results for $B \to PT$ decay modes are given in column III of the Tables III, IV, V(a) and V(b) and for $B \to VT$ decay modes are given in column III of the Tables VI, VII, VIII(a) and VIII(b) for various possible modes. We make the following observations:

**I) For $B \to PT$ meson decays:**

1. **$B \to PT$ decays involving $b \to c$ transition**

   a) $\Delta b = 1, \Delta C = 1, \Delta S = 0$ mode:

   i. Calculated branching ratio $B(B^- \to \pi^- D_2^0) = 6.7 \times 10^{-4}$ agrees well with the experiment value [19] $(7.8 \pm 1.4) \times 10^{-4}$, and $B(\bar{B}^0 \to \pi^- D_2^+) = 6.1 \times 10^{-4}$, is well below the experimental upper limit $< 2.2 \times 10^{-3}$.

   ii. Branching ratios of other dominant modes are $B(B^- \to D^0 a_2^-) = 1.8 \times 10^{-4}$, $B(\bar{B}_s^0 \to \pi^- D_{s2}^+) = 7.1 \times 10^{-4}$, and $B(\bar{B}_s^0 \to D^0 K_2^0) = 1.1 \times 10^{-4}$. We hope that these values are within the reach of the furure experiments.

   iii. Decays $\bar{B}^0 \to D^0 a_2^0$ and $\bar{B}^0 \to D^0 f_2$ have branching ratios of the order of $10^{-5}$, since these involve color-suppressed spectator process. The branching ratio of $\bar{B}^0 \to D^0 f_2'$ decay is further suppressed due to the $f_2 - f_2'$ mixing being close to the ideal mixing.

   iv. Decays $\bar{B}^0 \to \pi^0 D_2^0 / \eta D_2^0 / \eta' D_2^0 / D^+ a_2^- / D_s^+ K_2^- / K^- D_{s2}^+$ and $\bar{B}_s^0 \to K^0 D_2^0 / D_s^+ a_2^-$ are forbidden in the present analysis due to the vanishing matrix element between the vacuum and tensor meson. However, these may occur through an annihilation mechanism. The decays $\bar{B}^0 \to \pi^0 D_2^0 / D^+ a_2^-$ may also occur through elastic final state interactions (FSIs).

   b) $\Delta b = 1, \Delta C = 0, \Delta S = -1$ mode:

   i. Dominant modes are found to have branching ratios: $B(B^- \to D_s^- D_2^0) = 6.8 \times 10^{-4}$, $B(B^- \to \eta_c K_2^-) = 1.4 \times 10^{-4}$, $B(\bar{B}^0 \to D_s^- D_2^+) = 6.4 \times 10^{-4}$, $B(\bar{B}^0 \to \eta_c \bar{K}_2^0) = 1.3 \times 10^{-4}$, $B(\bar{B}_s^0 \to D_s^- D_{s2}^+) = 7.7 \times 10^{-4}$ and $B(\bar{B}_s^0 \to \eta_c f_2') = 1.3 \times 10^{-4}$.

   ii. Decays $B^- \to D^0 D_{s2}^- / D_s^- D_2^0 / K^- \chi_{c2}(1P)$, $\bar{B}^0 \to D^+ D_{s2}^-$ $/ D_s^- D_2^+ / \bar{K}^0 \chi_{c2}(1P)$ and $\bar{B}_s^0 \to \pi^0 \chi_{c2} / \eta \chi_{c2} / \eta' \chi_{c2} / D^+ D_2^-$

$/D^0\overline{D}_2^0$ $/D_s^+D_{s2}^-/D^-D_2^+/\overline{D}^0D_2^0/\eta_c a_2^0$ are forbidden in our work. Penguin diagrams may cause $B^- \to D^0 D_{s2}^- / D_s^- D_2^0$ and $\overline{B}^0 \to D^+ D_{s2}^- / D_s^- D_2^+$ decays, however these are likely to remain suppressed as these decays require $c\overline{c}$ pair to be created.

c) $\underline{\Delta b = 1, \Delta C = 0, \Delta S = 0 \text{ mode}:}$

  i. For dominant decays, we predict $B(B^- \to D^- D_2^0) = 2.5 \times 10^{-5}$, $B(\overline{B}^0 \to D^- D_2^+) = 2.4 \times 10^{-5}$ and $B(\overline{B}_s^0 \to D^- D_{s2}^+) = 2.9 \times 10^{-5}$.

  ii. Decays $B^- \to D^0 D_2^- / \pi^- \chi_{c2}(1P)$, $\overline{B}^0 \to D^0 \overline{D}_2^0 / D_s^- D_{s2}^+$ $/D^+ D_2^- / \overline{D}^0 D_2^0 / D_s^+ D_{s2}^- / \pi^0 \chi_{c2}(1P) / \eta \chi_{c2}(1P) / \eta' \chi_{c2}(1P)$ and $\overline{B}_s^0 \to K^0 \chi_{c2} / D_s^+ D_2^-$ are forbidden in our analysis. Annihilation diagrams, elastic FSI and penguin diagrams may generate these decays to the naked charm mesons. However, decays emitting charmonium $\chi_{c2}(1P)$ remains forbidden in the ideal mixing limit.

d) $\underline{\Delta b = 1, \Delta C = 1, \Delta S = -1 \text{ mode}:}$

  i. Branching ratios of the dominant decays are $B(B^- \to K^- D_2^0) = 4.8 \times 10^{-5}$, $B(\overline{B}^0 \to K^- D_2^+) = 4.5 \times 10^{-5}$ and $B(\overline{B}_s^0 \to K^- D_{s2}^+) = 5.2 \times 10^{-5}$.

  ii. Decays $\overline{B}^0 \to \overline{K}^0 D_2^0 / D^+ K_2^-$ and $\overline{B}_s^0 \to \pi^- D_2^+ / \pi^0 D_2^0 / \eta D_2^0$ $/\eta' D_2^0 / D^+ a_2^- / D^0 a_2^0 / D_s^+ K_2^-$ are forbidden in our analysis. Annihilation diagrams do not contribute to these decays. However, these may acquire nonzero branching ratios through elastic FSI.

## 2. $B \to PT$ decays involving $b \to u$ transition

a) $\Delta b = 1, \Delta C = 0, \Delta S = 0$ mode:

   i. $B(B^- \to \pi^- f_2) = 7.1 \times 10^{-6}$ is in good agreement with the experimental value $(8.2 \pm 2.5) \times 10^{-6}$ and $B(\bar{B}^0 \to \pi^- a_2^+) = 1.3 \times 10^{-5}$ is well below the experimental upper limit $< 3.0 \times 10^{-4}$.

   ii. $B^- \to K^0 K_2^- / K^- K_2^0$, $\bar{B}^0 \to K^+ K_2^- / K^0 \bar{K}_2^0 / K^0 \bar{K}_2^0 / \bar{K}^0 K_2^0 / K^- K_2^+ / \pi^+ a_2^-$ and $\bar{B}_s^0 \to K^+ a_2^- / K^0 a_2^0 / K^0 f_2 / K^0 f_2'$ are forbidden in the present analysis. Annihilation process and FSIs may generate these decays.

   iii. $B^- \to \pi^- \bar{K}_2^0$ and $\bar{B}^0 \to \pi^+ K_2^-$ are also forbidden in the present analysis which may be generated through annihilation diagram or elastic FSI.

b) $\Delta b = 1, \Delta C = -1, \Delta S = -1$ mode:

   i. Branching ratios $B(B^- \to D_s^- a_2^0) = 2.0 \times 10^{-5}$, $B(B^- \to D_s^- f_2') = 2.2 \times 10^{-5}$, $B(\bar{B}^0 \to D_s^- a_2^+) = 3.8 \times 10^{-5}$ and $B(\bar{B}_s^0 \to D_s^- K_2^+) = 2.6 \times 10^{-5}$ have relatively large branching ratios.

   ii. Decays $B^- \to \pi^0 D_{s2}^- / \eta D_{s2}^- / \eta' D_{s2}^- / \bar{K}^0 D_2^- / K^- D_2^0 / D^- \bar{K}_2^0$, $\bar{B}^0 \to \bar{K}^0 D_2^0 / \pi^+ D_{s2}^-$ and $\bar{B}_s^0 \to K^+ D_{s2}^- / \pi^+ D_2^- / \pi^0 \bar{D}_2^0 / \eta \bar{D}_2^0 / \eta' \bar{D}_2^0 / \bar{D}^0 a_2^0 / D^- a_2^+$ are forbidden in the present analysis. Annihilation and FSIs may generate these decays.

c) $\Delta b = 1, \Delta C = -1, \Delta S = 0$ mode:

   i. Branching ratios of $B(\bar{B}^0 \to \bar{D}^0 f_2) = 3.6 \times 10^{-8}$ is smaller than the experimental value $(1.2 \pm 0.4) \times 10^{-4}$. It may be noted that $W$-annihilation and $W$-exchange diagrams may also contribute to the $B$ decays under consideration. Normally, such contributions are expected to be suppressed due to the helicity and color arguments. Including the factorizable contribution of such diagrams, the decay amplitude of $\bar{B}^0 \to \bar{D}^0 f_2$ get modified to (leaving aside the scale factor $\frac{G_F}{\sqrt{2}} V_{ub} V_{cd}^*$)

$$A(\bar{B}^0 \to \bar{D}^0 f_2) = \frac{1}{\sqrt{2}} a_2 f_D \cos\phi_T F^{B \to f_2}(m_D^2) + \frac{1}{\sqrt{2}} a_2 f_B \cos\phi_T F^{f_2 \to D}(m_B^2). \quad (32)$$

Using $f_B = 0.176$ GeV, we find that the experimental branching ratio $B(\bar{B}^0 \to \bar{D}^0 f_2)$ requires $F^{f_2 \to D}(m_B^2) = -9.99$ GeV. This in turn enhances the branching ratio for $B^- \to D^- f_2$ to $1.2 \times 10^{-4}$.

ii. Dominant decay is $B(\bar{B}^0 \to D^- a_2^+) = 1.2 \times 10^{-6}$ and next order dominant decays are $B(B^- \to D^- f_2) = 6.9 \times 10^{-7}$ $B(B^- \to D^- a_2^0) = 6.5 \times 10^{-7}$ and $B(\bar{B}_s^0 \to D^- K_2^+) = 8.3 \times 10^{-7}$.

iii. Decays $B^- \to K^0 D_{s2}^- / \pi^0 D_2^- / \pi^- \bar{D}_2^0 / \eta D_2^- / \eta' D_2^- / D_s^- K_2^0 / \eta_c D_2^-$, $\bar{B}^0 \to K^+ D_{s2}^- / \pi^+ D_2^- / \pi^0 \bar{D}_2^0 / \eta \bar{D}_2^0 / \eta' \bar{D}_2^0 / D_s^- K_2^+ / \eta_c D_2^-$ and $\bar{B}_s^0 \to K^0 \bar{D}_2^0$ are forbidden in the present analysis. Annihilation diagrams may generate these decays.

d) $\Delta b = 1, \Delta C = 0, \Delta S = -1$ mode:

i. $B(B^- \to K^- f_2) = 0.54 \times 10^{-6}$ is smaller than the experimental value $(1.3^{+0.4}_{-0.5}) \times 10^{-6}$. This decay mode is also likely to have contribution from the W-annihilation and W-exchange processes. Including the factorizable contribution of such diagrams, the decay amplitudes of $B^- \to K^- f_2$ get modified to (putting aside the scale factor $\frac{G_F}{\sqrt{2}} V_{ub} V_{us}^*$)

$$A(B^- \to K^- f_2) = \frac{1}{\sqrt{2}} a_1 f_K \cos\phi_T F^{B \to f_2}(m_K^2) + \frac{1}{\sqrt{2}} a_1 f_B \cos\phi_T F^{f_2 \to K}(m_B^2). \quad (33)$$

As it is not possible to evaluate the form factor $F^{f_2 \to K}$ at $m_B^2$ even in the phenomenological models, it is treated as a free parameter. Taking $f_B = 0.176$ GeV, we find that the experimental branching ratio $B(B^- \to K^- f_2) = (1.3^{+0.4}_{-0.5}) \times 10^{-6}$ requires $F^{f_2 \to K}(m_B^2) = -0.083$ GeV. This value in turn enhances the branching ratio for $B^- \to K^- f_2$ through the W-annihilation contibution to $1.3 \times 10^{-6}$.

ii. Branching ratios of $B(B^- \to \eta K_2^-) = 1.2 \times 10^{-8}$ is small than the experimental value $(9.1 \pm 3.0) \times 10^{-6}$. Similar to $B^- \to K^- f_2$ decay, this decay mode is also likely to have contribution from the W-annihilation and W-exchange processes. Including the factorizable contribution of such diagrams, the decay amplitudes of $B \to \eta K_2$ get modified to (leaving aside the scale factor $\frac{G_F}{\sqrt{2}} V_{ub} V_{us}^*$)

$$A(B^- \to \eta K_2^-) = \frac{1}{\sqrt{2}} a_2 f_\eta \sin\phi_P F^{B \to K_2}(m_\eta^2) +$$
$$\frac{1}{\sqrt{2}} a_2 f_B \sin\phi_P F^{K_2 \to \eta}(m_B^2)$$
$$A(\overline{B}^0 \to \eta \overline{K}_2^0) = \frac{1}{\sqrt{2}} a_2 f_\eta \sin\phi_P F^{B \to K_2}(m_\eta^2) +$$
$$\frac{1}{\sqrt{2}} a_2 f_B \sin\phi_P F^{K_2 \to \eta}(m_B^2). \qquad (34)$$

For $f_B = 0.176$ GeV, we find that the experimental branching ratio $B(B^- \to \eta K_2^-) = (9.1 \pm 3.0) \times 10^{-6}$ requires $F^{K_2 \to \eta}(m_B^2) = -3.03$ GeV. This in turn enhances the branching ratio for $\overline{B}^0 \to \eta \overline{K}_2^0$ to $8.1 \times 10^{-6}$, which is consistent with the experimental value $(9.6 \pm 2.1) \times 10^{-6}$.

iii. Decays $B^- \to \pi^- \overline{K}_2^0 / \overline{K}^0 a_2^-$, $\overline{B}^0 \to \pi^+ K_2^- / \overline{K}^0 a_2^0 / \overline{K}^0 f_2 / \overline{K}^0 f_2'$ and $\overline{B}_s^0 \to K^+ K_2^- / K^0 \overline{K}_2^0 / \pi^+ a_2^- / \pi^0 a_2^0 / \pi^- a_2^+ / \eta a_2^0 / \overline{K}^0 K_2^0 / \eta' a_2^0$ are forbidden in the present analysis. Annihilation and FSIs may generate these decays.

**II) For $B \to VT$ meson decays:**

**1. $B \to VT$ decays involving $b \to c$ transition**

a) $\underline{\Delta b =1, \Delta C =1, \Delta S = 0 \text{ mode}:}$

  i. In the present analysis, branching ratios $B(B^- \to \rho^- D_2^0) = 1.3\times10^{-3}$ and $B(\bar{B}^0 \to \rho^- D_2^+) = 1.2\times10^{-3}$, consistent with the experimental upper limit $<4.7\times10^{-3}$ and $<4.9\times10^{-3}$.

  ii. Dominant decay has branching ratio $B(\bar{B}_s^0 \to \rho^- D_{s2}^+) = 2.1\times10^{-3}$.

  iii. Due to the vanishing matrix element between the vacuum and tensor meson, the following decays: $\bar{B}^0 \to \rho^0 D_2^0 / \omega D_2^0 / \phi D_2^0 / D^{*+}a_2^- / D_s^{*+}K_2^- / K^{*-}D_{s2}^+$ and $\bar{B}_s^0 \to K^{*0}D_2^0 / D_s^{*+}a_2^-$ are forbidden in the present analysis. But, these may appear through annihilation diagrams. Like $B \to PT$ decays, here also, $\bar{B}^0 \to \rho^0 D_2^0 / D^{*+}a_2^-$ may occur through elastic final state interactions (FSIs).

b) $\underline{\Delta b =1, \Delta C = 0, \Delta S = -1 \text{ mode}:}$

  i. Branching ratios of dominant decays are $B(\bar{B}_s^0 \to D_s^{*-} D_{s2}^+) = 1.3\times10^{-3}$, $B(\bar{B}^0 \to D_s^{*-} D_2^+) = 1.1\times10^{-3}$ and $B(B^- \to D_s^{*-} D_2^+) = 1.1\times10^{-3}$.

  ii. Analogous to $B \to PT$, $B^- \to D^{*0}D_{s2}^- / K^{*-}\chi_{c2}(1P)$, $\bar{B}^0 \to D^{*+}D_{s2}^- / \bar{K}^{*0}\chi_{c2}(1P)$ and $\bar{B}_s^0 \to \rho^0 \chi_{c2} / \omega \chi_{c2} / \phi \chi_{c2} / D^{*+}D_2^- / D^{*0}\bar{D}_2^0 / D_s^{*+}D_{s2}^- / D^{*-}D_2^+ / \bar{D}^{*0}D_2^0 / \psi a_2^0$ decays are forbidden in this work. $B^- \to D^{*0}D_{s2}^- / D_s^{*-}D_2^0$ and $\bar{B}^0 \to D^{*+}D_{s2}^- / D_s^{*-}D_2^+$ decays may get contribution through penguin diagrams. However requirement of $c\bar{c}$ pair creation may suppress these decays.

c) $\Delta b = 1, \Delta C = 0, \Delta S = 0$ mode:

   i. Here also, dominant decays have the branching ratios of the order of $10^{-5}$ except the decay $\bar{B}^0 \to \psi f_2'$ whose branching ratios comes out to be $2.0 \times 10^{-7}$.

   ii. Forbidden decays in this mode are $B^- \to D^{*0} D_2^- / \rho^- \chi_{c2}(1P)$, $\bar{B}^0 \to D^{*0} \bar{D}_2^0 / D_s^{*-} D_{s2}^+ / D^{*+} D_2^- / \bar{D}^{*0} D_2^0 / D_s^{*+} D_{s2}^- / \rho^0 \chi_{c2}(1P) / \omega \chi_{c2}(1P) / \phi \chi_{c2}(1P)$ and $\bar{B}_s^0 \to K^{*0} \chi_{c2} / D_s^{*+} D_2^-$. Annihilation diagrams, elastic FSI and penguin diagrams may generate these decays emitting naked charm mesons. However, decays emitting charmonium $\chi_{c2}(1P)$ remain forbidden in the ideal mixing limit.

d) $\Delta b = 1, \Delta C = 1, \Delta S = -1$ mode:

   i. The only dominant decay has branching ratio $B(\bar{B}_s^0 \to K^{*-} D_{s2}^+) = 1.1 \times 10^{-4}$.

   ii. In this mode $\bar{B}^0 \to \bar{K}^{*0} D_2^0 / D^{*+} K_2^-$ and $\bar{B}_s^0 \to \rho^- D_2^+ / \rho^0 D_2^0 / \omega D_2^0 / \phi D_2^0 / D^{*+} a_2^- / D^{*0} a_2^0 / D_s^{*+} K_2^-$ decays are forbidden in our analysis. These decays do not acquire contribution from annihilation process. However, elastic FSI may generate nonzero branching ratios for these decays.

2. $B \to VT$ decays involving $b \to u$ transition

   a) $\Delta b = 1, \Delta C = 0, \Delta S = 0$ mode:

   i. $B(B^- \to \rho^0 a_2^-) = 1.1 \times 10^{-6}$ is well below the experimental upper limit $< 7.2 \times 10^{-4}$.

   ii. Branching ratios $B(B^- \to \rho^- a_2^0) = 19.4 \times 10^{-6}$ and $B(\bar{B}^0 \to \rho^- a_2^+) = 36.2 \times 10^{-6}$ match well with the numerical values predicted by the J. Muñoz and N. Quintero [24].

   iii. In the present analysis $B^- \to K^{*0} K_2^- / K^{*-} K_2^0$, $\bar{B}^0 \to K^{*+} K_2^- / K^{*0} \bar{K}_2^0 / \bar{K}^{*0} K_2^0 / K^{*-} K_2^+ / \rho^+ a_2^-$ and $\bar{B}_s^0 \to K^{*+} a_2^- /$

$K^{*0}a_2^0 / K^{*0}f_2 / K^{*0}f_2'$ decays are forbidden. But these may get contribution from annihilation process and FSIs.

iv. Decays $B^- \to \rho^- \bar{K}_2^0$ and $\bar{B}^0 \to \rho^+ K_2^-$, which may be generated through annihilation diagram or elastic FSI, are also forbidden in the present analysis.

b) $\Delta b = 1, \Delta C = -1, \Delta S = -1$ mode:

  i. Obtained branching ratio $B(\bar{B}^0 \to D_s^{*-} a_2^+) = 3.7 \times 10^{-5}$ is well below the experimental upper limit $< 2.0 \times 10^{-4}$.

  ii. Also in this mode, $B^- \to \rho^0 D_{s2}^- / \omega D_{s2}^- / \phi D_{s2}^- / \bar{K}^{*0} D_2^- / K^{*-} D_2^0 / D^{*-}\bar{K}_2^0$, $\bar{B}^0 \to \bar{K}^{*0} D_2^0 / \rho^+ D_{s2}^-$ and $\bar{B}_s^0 \to K^{*+} D_{s2}^- / \rho^+ D_2^- / \rho^0 \bar{D}_2^0 / \omega \bar{D}_2^0 / \phi \bar{D}_2^0 / \bar{D}^{*0} a_2^0 / D^{*-} a_2^+$ decays are forbidden. Though, these may appear through annihilation and FSIs.

c) $\Delta b = 1, \Delta C = -1, \Delta S = 0$ mode:

  i. Dominant decays are $B(B^- \to D^{*-} a_2^+) = 1.8 \times 10^{-6}$, $B(B^- \to D^{*-} f_2) = 1.8 \times 10^{-6}$ and $B(\bar{B}_s^0 \to D^{*-} K_2^+) = 1.0 \times 10^{-6}$.

  ii. Here also, $B^- \to K^{*0} D_{s2}^- / \rho^0 D_2^- / \rho^- \bar{D}_2^0 / \omega D_2^- / \phi D_2^- / D_s^{*-} K_2^0$, $\bar{B}^0 \to K^{*+} D_{s2}^- / \rho^+ D_2^- / \rho^0 \bar{D}_2^0 / \omega \bar{D}_2^0 / \phi \bar{D}_2^0 / D_s^{*-} K_2^+$ and $\bar{B}_s^0 \to K^{*0} \bar{D}_2^0 / K^{*+} D_2^-$ decays are forbidden. Annihilation diagrams may generate these decays.

d) $\Delta b = 1, \Delta C = 0, \Delta S = -1$ mode:

  i. Branching ratios of $B(B^- \to \rho^0 K_2^-) = 0.006 \times 10^{-6}$, $B(B^- \to \phi K_2^-) = 0.006 \times 10^{-6}$ and $B(\bar{B}^0 \to \rho^0 \bar{K}_2^0) = 0.05 \times 10^{-6}$ are well below the experimental upper limit $< 1.5 \times 10^{-3}$, $< 3.4 \times 10^{-3}$ and $< 1.1 \times 10^{-3}$. $B(\bar{B}^0 \to \phi \bar{K}_2^0) = 0.05 \times 10^{-6}$ is well below the only experimentally observed value $(7.8 \pm 1.3) \times 10^{-6}$.

ii. Decays $B^- \to \rho^- \bar{K}_2^0 / \bar{K}^{*0} a_2^-$, $\bar{B}^0 \to \rho^+ K_2^- / \bar{K}^{*0} a_2^0 / \bar{K}^{*0} f_2 / \bar{K}^{*0} f_2'$ and $\bar{B}_s^0 \to \bar{K}^{*0} K_2^0 / K^{*0} \bar{K}_2^0 / \rho^+ a_2^- / \rho^0 a_2^0 / \rho^- a_2^+ / \omega a_2^0 / \phi a_2^0$ are forbidden in the present analysis. These may appear via annihilation and FSIs.

## V. SUMMARY AND CONCLUSIONS

In this paper, we have studied hadronic weak decays of bottom mesons emitting pseudoscalar/vector and tensor mesons. The matrix elements $\langle T(q_\mu)|J_\mu|0\rangle$ vanish due to the tracelessness of the polarization tensor $\in_{\mu\upsilon}$ of spin 2 meson and the auxiliary condition $q^\mu \in_{\mu\upsilon} = 0$. Therefore, either color-favored or color-suppressed diagrams contribute. Therefore, the analysis of these decays is free of constructive or destructive interference for color-favored and color-suppressed diagrams. We employ ISGW II model [3] to determine the $B \to T$ form factors appearing in the decay matrix element of weak currents involving $b \to c$ and $b \to u$ transitions. Consequently, we have obtained the decay amplitudes and calculated the branching ratios of $B \to PT/VT$ decays in CKM-favored and CKM-suppressed modes. We make the following conclusions:

### (i) $B \to PT$ decays:

Decays involving $b \to c$ transition have larger branching ratios of the order of $10^{-4}$ to $10^{-8}$ and decays involving $b \to u$ transition have branching ratios of the order of $10^{-5}$ to $10^{-11}$. Dominant decay modes involving $b \to c$ transition are $B(B^- \to D_s^- D_2^0) = 6.8 \times 10^{-4}$, $B(B^- \to \pi^- D_2^0) = 6.7 \times 10^{-4}$, $B(\bar{B}^0 \to D_s^- D_2^+) = 6.4 \times 10^{-4}$, $B(\bar{B}^0 \to \pi^- D_2^+) = 6.1 \times 10^{-4}$, $B(B^- \to D^0 a_2^-) = 1.8 \times 10^{-4}$, $B(B^- \to \eta_c K_2^-) = 1.4 \times 10^{-4}$, $B(\bar{B}^0 \to \eta_c \bar{K}_2^0) = 1.3 \times 10^{-4}$, $B(\bar{B}_s^0 \to D_s^- D_{s2}^+) = 7.7 \times 10^{-4}$, $B(\bar{B}_s^0 \to \pi^- D_{s2}^+) = 7.1 \times 10^{-4}$, $B(\bar{B}_s^0 \to \eta_c f_2') = 1.3 \times 10^{-4}$ and $B(\bar{B}_s^0 \to D^0 K_2^0) = 1.1 \times 10^{-4}$. Experimentally, the branching ratios of only five decay modes are measured and upper limits are available for six other decays. We find that the calculated branching ratio $B(B^- \to \pi^- f_2) = 7.1 \times 10^{-6}$ is in good agreement with the experimental value $(8.2 \pm 2.5) \times 10^{-6}$ whereas $B(B^- \to K^- f_2) = 5.4 \times 10^{-7}$ is smaller than the experimental value $(1.3^{+0.4}_{-0.5}) \times 10^{-6}$. $B$-decay requires contribution from $W$-annihilation diagram to bridge the gap between theoretical and experimental value.

### (ii) $B \to VT$ decays:

Here again, decays involving $b \to c$ transition have branching ratios ranging from $10^{-3}$ to $10^{-7}$, while decays involving $b \to u$ transition have branching ratios range from $10^{-5}$ to $10^{-11}$. Branching ratios of dominant decay modes are $B(\bar{B}_s^0 \to \rho^- D_{s2}^+) =$ $2.1\times10^{-3}$, $B(B^- \to \rho^- D_2^0) = 1.8\times10^{-3}$, $B(\bar{B}^0 \to \rho^- D_2^+) = 1.7\times10^{-3}$, $B(\bar{B}_s^0 \to D_s^{*-} D_{s2}^+) = 1.3\times10^{-3}$, $B(B^- \to D_s^{*-} D_2^0) = 1.1\times10^{-3}$ and $B(\bar{B}^0 \to D_s^{*-} D_2^+) = 1.1\times10^{-3}$. In contrast to the charm meson decays, the experimental data show constructive interference for $B$ meson decays involving both the color-favored and color-suppressed diagrams, giving $a_1 = 1.10 \pm 0.08$ and $a_2 = 0.20 \pm 0.02$. In the present analysis, the decay amplitude is proportional to only one QCD coefficient either $a_1$ (for color favored diagram) or $a_2$ (for color suppressed diagram), therefore our results remains unaffected from the interference pattern.

### (iii) Comparison with other models

We also compare our results with branching ratios calculated in the other models [17,23,24]. The predicted branching ratios in KLO [17] shown in 3rd column of tables III to VIII (b), while the prediction of MQ [24] are given in 4th column of tables V(a), V(b), VIII(a) and VIII(b). The prediction of KLO [17] are in general smaller as compared to the present branching ratios because of the difference in the form factors since different quark masses have been used in the two works.

MQ [24] have recently studied few charmless decays of $B \to PT$ and $B \to VT$ mode. Some of the branching ratios are smaller than our numerical value of branching ratios, while the others are large as compared to the present predictions, particularly for $\eta$ or $\eta'$ emitting decays. The disagreement with their predictions may be attributed to the difference in the form factors obtained in the covariant light-front approach (CLF) and inclusion of the non-factorizable contributions in their results. It may be noted that the form factors at small $q^2$ obtained in the CLF and ISGW II quark model agrees within 40% [3]. However, when $q^2$ increases $h(q^2)$, $b_+(q^2)$ and $b_-(q^2)$ increases more rapidly in the covariant light front model than in the ISGW II model. Another important fact is that the behavior of the form factor $k$ in both models is different, especially, for the decay $B \to \phi K_2$. The form factor $k(m_\phi^2)$ is bigger in ISGW II quark model than in light front quark model and decay constants used to calculate the numerical values are different in both the works.

Branching ratios have also been calculated by Cheng [25]. For $B \to PT$ decays his predictions $B(B^- \to \pi^- D_2^0) = 6.7\times10^{-4}$ and $B(\bar{B}^0 \to \pi^- D_2^+) = 6.1\times10^{-4}$ match well with the numerical branching ratios obtained in the present work. However, the other branching ratios $B(B^- \to D_s^- D_2^0) = 4.2\times10^{-4}$, $B(\bar{B}^0 \to D_s^- D_2^+) = 3.8\times10^{-4}$ and $B(\bar{B}_s^0 \to \pi^- D_{s2}^+) = 3.8\times10^{-4}$ are different from our results owing to the different values used for the decay constant $f_{D_s}$. For $B \to VT$ decays his predictions $B(B^- \to \rho^- D_2^0) = 1.8\times10^{-3}$, $B(B^- \to D_s^{*-} D_2^0) = 1.1\times10^{-3}$, $B(\bar{B}^0 \to \rho^- D_2^+) = 1.7\times10^{-3}$, $B(\bar{B}^0 \to D_s^{*-} D_2^+) = 1.0\times10^{-3}$ and $B(\bar{B}_s^0 \to \rho^- D_{s2}^+) = 2.1\times10^{-3}$ match well with the numerical branching ratios obtained in the present work.

The Belle collaboration is currently searching for some $B \to PT$ and $B \to VT$ modes and their preliminary results indicate that the branching ratios for these may not be very small compared to $B \to PP$ modes. We hope our predictions would be within the reach of the current experiments. Observation of these decays in the *B* experiments such as Belle, Babar, BTeV, LHC and so on will be crucial in testing the ISGW II and other quark models as well as validity of the factorization scheme.

**Acknowledgement**

One of the authors (N. S.) is thankful to the University Grant Commission, New Delhi, for the financial assistance.

**References**

1. M. Wirbel, B. Stech and M. Bauer, Z. Phys. C **29** (1985) 637; M. Bauer, B. Stech and M. Wirbel, Z. Phys. C **34** (1987) 103; M. Wirbel, Prog. Part. Nucl. Phys. **21** (1988) 33, and references therein.

2. N. Isgur, D. Scora, B. Grinstein and M.B. Wise, Phys. Rev. D **39** (1989) 799, and references therein.

3. D. Scora and N. Isgur, Phys. Rev. D **52** (1995) 2783, and references therein.

4. J. Bjorken, in New Developments in High-Energy Physics Proceddings of 4$^{th}$ Workshop, Crete, Greece,1988, edited by E.G. Floratos and A Verganelakis [Nucl. Phys. B (Proc. Suppl.) **11** (1989) 325]; J.M. Cline, W. F. Palmer and G. Kramer, Phys. Rev. D **40** (1989) 793; A. Deandrea *et. al.*, Phys. Lett. B **318** (1993) 549; W. Jaus, Phys. Rev. D **41** (1990) 3394; M. Tanimoto, K. Goda and K. Senba, *ibid*. **4**2 (1994) 3741; Fayyazuddin and Riazuddin, *ibid*. **49** (1994) 3385.

5. T. Mannel, W. Roberts and Z. Ryzak, Phys. Lett. B **248** (1990) 392; M.J. Dugan and B. Grinstein, *ibid*. **255** (1991) 583, and references therein; A Ali and T. Mannel, *ibid*. **264** (1991) 447; G. Kramer, T. Mannel and W.F. Palmer, Z. Phys. C **55** (1992) 497; A.F. Falk, M.B. Wise and I. Dunietz, Phys. Rev. D **51** (1995) 1183.

6. M. Neubert, V. Rieckert, B. Stech, and Q. P. Xu, in *Heavy Flavours*, edited by A.J. Buras and H. Lindner, World Scientific, Singapore, 1992 and references therein; S. Stone, talk presented at 5th Int. Symposium on *Heavy Flavor Physics*, Montrêal, July 6-10, 1993.

7. T.E. Browder and K. Honscheid, Prog. Part. Nucl. Phys. **35** (1995) 81, and references therein.

8. M. Gourdin, A.N. Kamal and X.Y. Pham, Phys. Rev. Lett. **73** (1994) 3197; A.N. Kamal and T.N. Pham, Phys. Rev. D **50** (1994) 395; M. Gourdin *et al.*, Phys. Lett. B **333** (1994) 507.

9. M.S. Alam *et al.*, Phys. Rev. D **50** (1994) 43.

10. S.M. Sheikholeslami and R.C. Verma, Int. J Mod. Phys. A **7** (1992).

11. G. Kramer and W.F. Palmer, Phys. Rev. D **46** (1992) 3197.

12. C.E. Carlson and J. Milana, Phys. Rev. D **49** (1994) 5908.

13. A.C. Katoch and R.C. Verma, Phys. Rev. D **52** (1995) 1717; **55** (1997) 7316 (E).

14. G. Lopez Castro and J.H. Munoz, Phys. Rev. D **55** (1997) 5581.

15. J.H. Munoz, A.A. Rojas and G. Lopez Castro, Phys. Rev. D **59** (1999) 077504.

16. C.S. Kim, B.H. Lim and Sechul Oh, Eur. Phys. J. C **22** (2002) 683; Eur. Phys. J. C **22** (2002) 683; C.S. Kim, J.P. Lee and Sechul Oh, hep-ph/0205262; Phys. Rev. D **67** (2003) 014002.

17. C.S. Kim, J.P. Lee and Sechul Oh, Phys. Rev. D **67** (2003) 014011.

18. C. Amsler *et al.* (Particle Data Group), Phys. Lett. B **667** (2008).

19. Belle Collaboration, K. Abe *et al.*, Phys. Rev. D **69** (2004) 112002.

20. Sechul Oh, Phys. Rev. D **60** (1999) 034006; M. Gronau and J.L. Rosner, *ibdi*. **61**, (2000) 073008.

21. D. Spehler and S.F. Novaes, Phys. Rev. D **44** (1991) 3990.

22. Rohit Dhir, Neelesh Sharma and R.C. Verma, J. Phys. G: Nucl. Part. Phys. **35** (2008) 085002.

23. C. H. Chen and C.Q. Geng, Phys. Rev. D **75** (2007) 054010.

24. J. Muñoz and N. Quintero, J. Phys. G: Nucl. Part. Phys. **36** (2009) 095004.

25. H.Y Cheng, Phys. Rev. D **68** (2003) 094005.

**Table I. The parameter $\beta$ for *s*-wave and *p*-wave mesons in the ISGW II model**

| Quark content | $u\bar{d}$ | $u\bar{s}$ | $s\bar{s}$ | $c\bar{u}$ | $c\bar{s}$ | $u\bar{b}$ | $s\bar{b}$ |
|---|---|---|---|---|---|---|---|
| $\beta_S$ (GeV) | 0.41 | 0.44 | 0.53 | 0.45 | 0.56 | 0.43 | 0.54 |
| $\beta_P$ (GeV) | 0.28 | 0.30 | 0.33 | 0.33 | 0.38 | 0.35 | 0.41 |

**Table II. Form factors of $B \to T$ transition at $q^2 = t_m$ in the ISGW II quark model**

| Transition | $k$ | $b_+$ | $b_-$ |
|---|---|---|---|
| $B \to a_2$ | 0.432 | -0.013 | 0.015 |
| $B \to f_2$ | 0.425 | -0.014 | 0.014 |
| $B \to K_2$ | 0.480 | -0.015 | 0.015 |
| $B \to D_2$ | 0.677 | -0.013 | 0.013 |
| $B_s \to f'_2$ | 0.572 | -0.016 | 0.017 |
| $B_s \to K_2$ | 0.492 | -0.013 | 0.015 |
| $B_s \to D_{s2}$ | 0.854 | -0.015 | 0.016 |

**Table III. Branching ratios of** $B \to PT$ **decays in CKM-favored mode involving** $b \to c$ **transition**

| Decay | Branching ratios | |
|---|---|---|
| | This Work | KLO [16] |
| $\Delta b = 1, \Delta C = 1, \Delta S = 0$ | | |
| $B^- \to \pi^- D_2^0$ | $6.7 \times 10^{-4}$ | $3.5 \times 10^{-4}$ |
| $B^- \to D^0 a_2^-$ | $1.8 \times 10^{-4}$ | $1.0 \times 10^{-4}$ |
| $\overline{B}^0 \to \pi^- D_2^+$ | $6.1 \times 10^{-4}$ | $3.3 \times 10^{-4}$ |
| $\overline{B}^0 \to D^0 a_2^0$ | $8.2 \times 10^{-5}$ | $4.8 \times 10^{-4}$ |
| $\overline{B}^0 \to D^0 f_2$ | $8.8 \times 10^{-5}$ | $5.3 \times 10^{-5}$ |
| $\overline{B}^0 \to D^0 f_2'$ | $1.7 \times 10^{-6}$ | $0.62 \times 10^{-6}$ |
| $\overline{B}_s^0 \to \pi^- D_{s2}^+$ | $7.1 \times 10^{-4}$ | - |
| $\overline{B}_s^0 \to D^0 K_2^0$ | $1.1 \times 10^{-4}$ | - |
| $\Delta b = 1, \Delta C = 0, \Delta S = -1$ | | |
| $B^- \to D_s^- D_2^0$ | $6.8 \times 10^{-4}$ | $4.9 \times 10^{-4}$ |
| $B^- \to \eta_c K_2^-$ | $1.4 \times 10^{-4}$ | $1.1 \times 10^{-4}$ |
| $\overline{B}^0 \to D_s^- D_2^+$ | $6.4 \times 10^{-4}$ | $4.6 \times 10^{-4}$ |
| $\overline{B}^0 \to \eta_c \overline{K}_2^0$ | $1.3 \times 10^{-4}$ | $9.6 \times 10^{-5}$ |
| $\overline{B}_s^0 \to D_s^- D_{s2}^-$ | $7.7 \times 10^{-4}$ | - |
| $\overline{B}_s^0 \to \eta_c f_2$ | $2.7 \times 10^{-6}$ | - |
| $\overline{B}_s^0 \to \eta_c f_2'$ | $1.3 \times 10^{-4}$ | - |

**Table IV. Branching ratios of** $B \to PT$ **decays in CKM-suppressed mode involving** $b \to c$ **transition**

| Decay | Branching ratios | |
|---|---|---|
| | **This Work** | **KLO [16]** |
| $\Delta b = 1, \Delta C = 1, \Delta S = -1$ | | |
| $B^- \to K^- D_2^0$ | $4.8 \times 10^{-5}$ | $2.5 \times 10^{-5}$ |
| $B^- \to D^0 K_2^-$ | $8.7 \times 10^{-6}$ | $7.3 \times 10^{-6}$ |
| $\overline{B}^0 \to K^- D_2^+$ | $4.5 \times 10^{-5}$ | $2.4 \times 10^{-5}$ |
| $\overline{B}^0 \to D^0 \overline{K}_2^0$ | $8.1 \times 10^{-6}$ | $6.8 \times 10^{-6}$ |
| $\overline{B}_s^0 \to K^- D_{s2}^+$ | $5.2 \times 10^{-5}$ | - |
| $\overline{B}_s^0 \to D^0 f_2$ | $9.9 \times 10^{-8}$ | - |
| $\overline{B}_s^0 \to D^0 f_2'$ | $6.7 \times 10^{-6}$ | - |
| $\Delta b = 1, \Delta C = 0, \Delta S = 0$ | | |
| $B^- \to D^- D_2^0$ | $2.5 \times 10^{-5}$ | $2.2 \times 10^{-5}$ |
| $B^- \to \eta_c a_2^-$ | $9.2 \times 10^{-6}$ | $4.9 \times 10^{-6}$ |
| $\overline{B}^0 \to D^- D_2^+$ | $2.4 \times 10^{-5}$ | $2.1 \times 10^{-5}$ |
| $\overline{B}^0 \to \eta_c a_2^0$ | $4.3 \times 10^{-6}$ | $2.3 \times 10^{-6}$ |
| $\overline{B}^0 \to \eta_c f_2$ | $4.8 \times 10^{-6}$ | $2.7 \times 10^{-6}$ |
| $\overline{B}^0 \to \eta_c f_2'$ | $6.7 \times 10^{-8}$ | $0.02 \times 10^{-6}$ |
| $\overline{B}_s^0 \to D^- D_{s2}^+$ | $2.9 \times 10^{-5}$ | - |
| $\overline{B}_s^0 \to \eta_c K_2^0$ | $6.9 \times 10^{-6}$ | - |

**Table V (a). Branching ratios of** $B \to PT$ **decays involving** $b \to u$ **transition**

| Decays | Branching ratios | | |
|---|---|---|---|
| | This Work | KLO [16] | MQ [24] |
| $\Delta b = 1, \Delta C = -1, \Delta S = -1$ | | | |
| $B^- \to \overline{D}^0 K_2^-$ | $1.3 \times 10^{-6}$ | $1.2 \times 10^{-6}$ | - |
| $B^- \to D_s^- a_2^0$ | $2.0 \times 10^{-5}$ | $9.4 \times 10^{-6}$ | - |
| $B^- \to D_s^- f_2$ | $2.2 \times 10^{-5}$ | $1. \times 10^{-5}$ | - |
| $B^- \to D_s^- f_2'$ | $4.3 \times 10^{-7}$ | $0.12 \times 10^{-6}$ | - |
| $\overline{B}^0 \to \overline{D}^0 \overline{K}_2^0$ | $1.2 \times 10^{-6}$ | $1.1 \times 10^{-6}$ | - |
| $\overline{B}^0 \to D_s^- a_2^+$ | $3.8 \times 10^{-5}$ | $1.8 \times 10^{-5}$ | - |
| $\overline{B}_s^0 \to D_s^- K_2^+$ | $2.6 \times 10^{-5}$ | - | - |
| $\overline{B}_s^0 \to \overline{D}^0 f_2$ | $1.5 \times 10^{-8}$ | - | - |
| $\overline{B}_s^0 \to \overline{D}^0 f_2'$ | $1.0 \times 10^{-6}$ | - | - |
| $\Delta b = 1, \Delta C = 0, \Delta S = 0$ | | | |
| $B^- \to \pi^- a_2^0$ | $6.7 \times 10^{-6}$ | $2.6 \times 10^{-6}$ | $4.38 \times 10^{-6}$ |
| $B^- \to \pi^- f_2$ | $7.1 \times 10^{-6}$ | - | - |
| $B^- \to \pi^- f_2'$ | $1.5 \times 10^{-7}$ | - | - |
| $B^- \to \pi^0 a_2^-$ | $0.38 \times 10^{-6}$ | $0.001 \times 10^{-6}$ | $0.015 \times 10^{-6}$ |
| $B^- \to \eta a_2^-$ | $0.23 \times 10^{-6}$ | $0.29 \times 10^{-6}$ | $45.8 \times 10^{-6}$ |
| $B^- \to \eta' a_2^-$ | $0.13 \times 10^{-6}$ | $1.31 \times 10^{-6}$ | $71.3 \times 10^{-6}$ |
| $\overline{B}^0 \to \pi^- a_2^+$ | $13.0 \times 10^{-6}$ | $4.88 \times 10^{-6}$ | $8.19 \times 10^{-6}$ |
| $\overline{B}^0 \to \pi^0 a_2^0$ | $0.18 \times 10^{-6}$ | $0.0003 \times 10^{-6}$ | $0.007 \times 10^{-6}$ |
| $\overline{B}^0 \to \pi^0 f_2$ | $1.9 \times 10^{-7}$ | - | - |
| $\overline{B}^0 \to \pi^0 f_2'$ | $3.9 \times 10^{-9}$ | - | - |
| $\overline{B}^0 \to \eta a_2^0$ | $0.11 \times 10^{-6}$ | $0.14 \times 10^{-6}$ | $25.2 \times 10^{-6}$ |
| $\overline{B}^0 \to \eta f_2$ | $1.1 \times 10^{-7}$ | - | - |
| $\overline{B}^0 \to \eta f_2'$ | $2.4 \times 10^{-9}$ | - | - |
| $\overline{B}^0 \to \eta' a_2^0$ | $0.06 \times 10^{-6}$ | $0.62 \times 10^{-6}$ | $43.3 \times 10^{-6}$ |
| $\overline{B}^0 \to \eta' f_2$ | $6.3 \times 10^{-8}$ | - | - |
| $\overline{B}^0 \to \eta' f_2'$ | $1.3 \times 10^{-9}$ | - | - |
| $\overline{B}_s^0 \to \pi^- K_2^+$ | $7.8 \times 10^{-6}$ | - | - |
| $\overline{B}_s^0 \to \pi^0 K_2^0$ | $2.2 \times 10^{-7}$ | - | - |
| $\overline{B}_s^0 \to \eta K_2^0$ | $1.3 \times 10^{-7}$ | - | - |
| $\overline{B}_s^0 \to \eta' K_2^0$ | $7.5 \times 10^{-8}$ | - | - |

**Table V (b). Branching ratios of $B \to PT$ decays involving $b \to u$ transition**

| Decays | Branching ratios | | |
|---|---|---|---|
| | **This Work** | **KLO [16]** | **MQ [24]** |
| $\Delta b = 1, \Delta C = 0, \Delta S = -1$ | | | |
| $B^- \to K^- a_2^0$ | $0.51 \times 10^{-6}$ | $0.31 \times 10^{-6}$ | $0.39 \times 10^{-6}$ |
| $B^- \to K^- f_2$ | $5.4 \times 10^{-7}$ | - | - |
| $B^- \to K^- f_2'$ | $1.5 \times 10^{-8}$ | - | - |
| $B^- \to \pi^0 K_2^-$ | $0.02 \times 10^{-6}$ | $0.09 \times 10^{-6}$ | $0.15 \times 10^{-6}$ |
| $B^- \to \eta K_2^-$ | $0.01 \times 10^{-6}$ | $0.03 \times 10^{-6}$ | $1.19 \times 10^{-6}$ |
| $B^- \to \eta' K_2^-$ | $0.007 \times 10^{-6}$ | $1.40 \times 10^{-6}$ | $2.70 \times 10^{-6}$ |
| $\bar{B}^0 \to K^- a_2^+$ | $0.95 \times 10^{-6}$ | $0.58 \times 10^{-6}$ | $0.73 \times 10^{-6}$ |
| $\bar{B}^0 \to \pi^0 \bar{K}_2^0$ | $0.02 \times 10^{-6}$ | $0.08 \times 10^{-6}$ | $0.13 \times 10^{-6}$ |
| $\bar{B}^0 \to \eta \bar{K}_2^0$ | $0.01 \times 10^{-6}$ | $0.03 \times 10^{-6}$ | $1.09 \times 10^{-6}$ |
| $\bar{B}^0 \to \eta' \bar{K}_2^0$ | $0.006 \times 10^{-6}$ | $1.3 \times 10^{-6}$ | $2.46 \times 10^{-6}$ |
| $\bar{B}_s^0 \to K^- K_2^+$ | $5.9 \times 10^{-7}$ | - | - |
| $\bar{B}_s^0 \to \pi^0 f_2$ | $1.9 \times 10^{-10}$ | - | - |
| $\bar{B}_s^0 \to \pi^0 f_2'$ | $1.4 \times 10^{-8}$ | - | - |
| $\bar{B}_s^0 \to \eta f_2$ | $1.1 \times 10^{-10}$ | - | - |
| $\bar{B}_s^0 \to \eta f_2'$ | $8.3 \times 10^{-9}$ | - | - |
| $\bar{B}_s^0 \to \eta' f_2$ | $6.5 \times 10^{-11}$ | - | - |
| $\bar{B}_s^0 \to \eta' f_2'$ | $4.7 \times 10^{-9}$ | - | - |
| $\Delta b = 1, \Delta C = -1, \Delta S = 0$ | | | |
| $B^- \to D^- a_2^0$ | $6.5 \times 10^{-7}$ | - | - |
| $B^- \to D^- f_2$ | $6.9 \times 10^{-7}$ | - | - |
| $B^- \to D^- f_2'$ | $1.4 \times 10^{-7}$ | - | - |
| $B^- \to \bar{D}^0 a_2^-$ | $7.3 \times 10^{-8}$ | - | - |
| $\bar{B}^0 \to D^- a_2^+$ | $1.2 \times 10^{-6}$ | - | - |
| $\bar{B}^0 \to \bar{D}^0 a_2^0$ | $3.4 \times 10^{-8}$ | - | - |
| $\bar{B}^0 \to \bar{D}^0 f_2$ | $3.6 \times 10^{-8}$ | - | - |
| $\bar{B}^0 \to \bar{D}^0 f_2'$ | $7.1 \times 10^{-10}$ | - | - |
| $\bar{B}_s^0 \to D^- K_2^+$ | $8.3 \times 10^{-7}$ | - | - |
| $\bar{B}_s^0 \to \bar{D}^0 K_2^0$ | $4.6 \times 10^{-8}$ | - | - |

**Table VI. Branching ratios of $B \to VT$ decays in CKM-favored mode involving $b \to c$ transition**

| Decay | Branching ratios | |
|---|---|---|
| | **This Work** | **KLO [16]** |
| $\Delta b = 1, \Delta C = 1, \Delta S = 0$ | | |
| $B^- \to \rho^- D_2^0$ | $1.9 \times 10^{-3}$ | $1.0 \times 10^{-3}$ |
| $B^- \to D^{*0} a_2^-$ | $2.6 \times 10^{-4}$ | $1.7 \times 10^{-4}$ |
| $\bar{B}^0 \to \rho^- D_2^+$ | $1.8 \times 10^{-3}$ | $0.92 \times 10^{-3}$ |
| $\bar{B}^0 \to D^{*0} a_2^0$ | $1.2 \times 10^{-4}$ | $0.78 \times 10^{-4}$ |
| $\bar{B}^0 \to D^{*0} f_2$ | $1.3 \times 10^{-4}$ | $0.84 \times 10^{-4}$ |
| $\bar{B}^0 \to D^{*0} f_2'$ | $2.6 \times 10^{-6}$ | $1.1 \times 10^{-6}$ |
| $\bar{B}_s^0 \to \rho^- D_{s2}^+$ | $2.1 \times 10^{-3}$ | - |
| $\bar{B}_s^0 \to D^{*0} K_2^0$ | $1.7 \times 10^{-4}$ | - |
| $\Delta b = 1, \Delta C = 0, \Delta S = -1$ | | |
| $B^- \to D_s^{*-} D_2^0$ | $1.1 \times 10^{-3}$ | $1.2 \times 10^{-3}$ |
| $B^- \to \psi K_2^-$ | $4.6 \times 10^{-4}$ | $3.8 \times 10^{-4}$ |
| $\bar{B}^0 \to D_s^{*-} D_2^+$ | $1.1 \times 10^{-4}$ | $1.1 \times 10^{-3}$ |
| $\bar{B}^0 \to \psi \bar{K}_2^0$ | $4.3 \times 10^{-4}$ | $3.5 \times 10^{-4}$ |
| $\bar{B}_s^0 \to D_s^{*-} D_{s2}^+$ | $1.3 \times 10^{-3}$ | - |
| $\bar{B}_s^0 \to \psi f_2$ | $8.3 \times 10^{-6}$ | - |
| $\bar{B}_s^0 \to \psi f_2'$ | $4.3 \times 10^{-4}$ | - |

**Table VII. Branching ratios of** $B \to VT$ **decays in CKM-suppressed mode involving** $b \to c$ **transition**

| Decay | Branching ratios | |
|---|---|---|
| | This Work | KLO [16] |
| $\Delta b = 1, \Delta C = 1, \Delta S = -1$ | | |
| $B^- \to K^{*-} D_2^0$ | $9.7 \times 10^{-5}$ | $5.2 \times 10^{-5}$ |
| $B^- \to D^{*0} K_2^-$ | $1.4 \times 10^{-5}$ | $1.2 \times 10^{-5}$ |
| $\bar{B}^0 \to K^{*-} D_2^+$ | $9.1 \times 10^{-5}$ | $4.9 \times 10^{-5}$ |
| $\bar{B}^0 \to D^{*0} \bar{K}_2^0$ | $1.3 \times 10^{-5}$ | $1.1 \times 10^{-5}$ |
| $\bar{B}_s^0 \to K^{*-} D_{s2}^+$ | $1.1 \times 10^{-4}$ | - |
| $\bar{B}_s^0 \to D^{*0} f_2$ | $1.6 \times 10^{-7}$ | - |
| $\bar{B}_s^0 \to D^{*0} f_2'$ | $1.1 \times 10^{-5}$ | - |
| $\Delta b = 1, \Delta C = 0, \Delta S = 0$ | | |
| $B^- \to D^{*-} D_2^0$ | $6.1 \times 10^{-5}$ | $5.3 \times 10^{-5}$ |
| $B^- \to \psi a_2^-$ | $2.6 \times 10^{-5}$ | $1.6 \times 10^{-5}$ |
| $\bar{B}^0 \to D^{*-} D_2^+$ | $5.7 \times 10^{-5}$ | $5.0 \times 10^{-5}$ |
| $\bar{B}^0 \to \psi a_2^0$ | $12.3 \times 10^{-6}$ | $7.7 \times 10^{-6}$ |
| $\bar{B}^0 \to \psi f_2$ | $13.3 \times 10^{-6}$ | $8.4 \times 10^{-6}$ |
| $\bar{B}^0 \to \psi f_2'$ | $0.2 \times 10^{-6}$ | $0.09 \times 10^{-6}$ |
| $\bar{B}_s^0 \to D^{*-} D_{s2}^+$ | $1.2 \times 10^{-5}$ | - |
| $\bar{B}_s^0 \to \psi K_2^0$ | $2.1 \times 10^{-5}$ | - |

**Table VIII (a). Branching ratios of $B \to VT$ decays involving $b \to u$ transition**

| Decays | Branching ratios | | |
|---|---|---|---|
| | This Work | KLO [16] | MQ [24] |
| $\Delta b = 1, \Delta C = -1, \Delta S = -1$ | | | |
| $B^- \to \bar{D}^{*0} K_2^-$ | $2.1 \times 10^{-6}$ | $1.9 \times 10^{-6}$ | - |
| $B^- \to D_s^{*-} a_2^0$ | $19.6 \times 10^{-6}$ | $15.5 \times 10^{-6}$ | - |
| $B^- \to D_s^{*-} f_2$ | $20.7 \times 10^{-6}$ | $16.7 \times 10^{-6}$ | - |
| $B^- \to D_s^{*-} f_2'$ | $0.4 \times 10^{-6}$ | $0.2 \times 10^{-6}$ | - |
| $\bar{B}^0 \to \bar{D}^{*0} \bar{K}_2^0$ | $1.9 \times 10^{-6}$ | $1.8 \times 10^{-6}$ | - |
| $\bar{B}^0 \to D_s^{*-} a_2^+$ | $36.7 \times 10^{-6}$ | $29.1 \times 10^{-6}$ | - |
| $\bar{B}_s^0 \to D_s^{*-} K_2^+$ | $26.1 \times 10^{-6}$ | - | - |
| $\bar{B}_s^0 \to \bar{D}^{*0} f_2$ | $2.4 \times 10^{-8}$ | - | - |
| $\bar{B}_s^0 \to \bar{D}^{*0} f_2'$ | $1.6 \times 10^{-6}$ | - | - |
| $\Delta b = 1, \Delta C = 0, \Delta S = 0$ | | | |
| $B^- \to \rho^0 a_2^-$ | $1.1 \times 10^{-6}$ | $0.007 \times 10^{-6}$ | $0.07 \times 10^{-6}$ |
| $B^- \to \rho^- a_2^0$ | $19.4 \times 10^{-6}$ | $7.3 \times 10^{-6}$ | $19.4 \times 10^{-6}$ |
| $B^- \to \rho^- f_2$ | $20.4 \times 10^{-6}$ | - | - |
| $B^- \to \rho^- f_2'$ | $4.4 \times 10^{-7}$ | - | - |
| $B^- \to \omega a_2^-$ | $0.07 \times 10^{-6}$ | $0.01 \times 10^{-6}$ | $0.14 \times 10^{-6}$ |
| $B^- \to \phi a_2^-$ | $1.1 \times 10^{-6}$ | $0.004 \times 10^{-6}$ | $0.019 \times 10^{-6}$ |
| $\bar{B}^0 \to \rho^- a_2^+$ | $36.2 \times 10^{-6}$ | $14.7 \times 10^{-6}$ | $36.2 \times 10^{-6}$ |
| $\bar{B}^0 \to \rho^0 a_2^0$ | $0.5 \times 10^{-6}$ | $0.003 \times 10^{-6}$ | $0.03 \times 10^{-6}$ |
| $\bar{B}^0 \to \rho^0 f_2$ | $5.7 \times 10^{-7}$ | - | - |
| $\bar{B}^0 \to \rho^0 f_2'$ | $1.1 \times 10^{-8}$ | - | - |
| $\bar{B}^0 \to \omega a_2^0$ | $0.03 \times 10^{-6}$ | $0.005 \times 10^{-6}$ | $0.07 \times 10^{-6}$ |
| $\bar{B}^0 \to \omega f_2$ | $3.4 \times 10^{-8}$ | - | - |
| $\bar{B}^0 \to \omega f_2'$ | $7.2 \times 10^{-10}$ | - | - |
| $\bar{B}^0 \to \phi a_2^0$ | $0.5 \times 10^{-6}$ | $0.002 \times 10^{-6}$ | $0.009 \times 10^{-6}$ |
| $\bar{B}^0 \to \phi f_2$ | $5.3 \times 10^{-7}$ | - | - |
| $\bar{B}^0 \to \phi f_2'$ | $1.1 \times 10^{-8}$ | - | - |
| $\bar{B}_s^0 \to \rho^- K_2^+$ | $2.3\times10^{-5}$ | - | - |
| $\bar{B}_s^0 \to \rho^0 K_2^0$ | $6.4\times10^{-7}$ | - | - |
| $\bar{B}_s^0 \to \omega K_2^0$ | $4.0\times10^{-8}$ | - | - |
| $\bar{B}_s^0 \to \phi K_2^0$ | $6.4\times10^{-7}$ | - | - |

**Table VIII (b). Branching ratios of $B \to VT$ decays involving $b \to u$ transition**

| Decays | Branching ratios | | |
|---|---|---|---|
| | This Work | KLO [16] | MQ [24] |
| $\Delta b = 1, \Delta C = 0, \Delta S = -1$ | | | |
| $B^- \to K^{*-} a_2^0$ | $1.0 \times 10^{-6}$ | $1.9 \times 10^{-6}$ | $2.8 \times 10^{-6}$ |
| $B^- \to K^{*-} f_2$ | $1.1 \times 10^{-6}$ | - | - |
| $B^- \to K^{*-} f_2'$ | $2.3 \times 10^{-8}$ | - | - |
| $B^- \to \rho^0 K_2^-$ | $0.06 \times 10^{-6}$ | $0.253 \times 10^{-6}$ | $0.74 \times 10^{-6}$ |
| $B^- \to \omega K_2^-$ | $0.004 \times 10^{-6}$ | $0.112 \times 10^{-6}$ | $0.06 \times 10^{-6}$ |
| $B^- \to \phi K_2^-$ | $0.06 \times 10^{-6}$ | $2.2 \times 10^{-6}$ | $9.2 \times 10^{-6}$ |
| $\bar{B}^0 \to K^{*-} a_2^+$ | $1.9 \times 10^{-6}$ | $3.5 \times 10^{-6}$ | $7.3 \times 10^{-6}$ |
| $\bar{B}^0 \to \rho^0 \bar{K}_2^0$ | $0.05 \times 10^{-6}$ | $0.235 \times 10^{-6}$ | $0.68 \times 10^{-6}$ |
| $\bar{B}^0 \to \omega \bar{K}_2^0$ | $0.003 \times 10^{-6}$ | $0.104 \times 10^{-6}$ | $0.053 \times 10^{-6}$ |
| $\bar{B}^0 \to \phi \bar{K}_2^0$ | $0.05 \times 10^{-6}$ | $2.0 \times 10^{-6}$ | $8.5 \times 10^{-6}$ |
| $\bar{B}_s^0 \to K^{*-} K_2^+$ | $1.2 \times 10^{-6}$ | - | - |
| $\bar{B}_s^0 \to \rho^0 f_2$ | $5.6 \times 10^{-10}$ | - | - |
| $\bar{B}_s^0 \to \rho^0 f_2'$ | $4.1 \times 10^{-8}$ | - | - |
| $\bar{B}_s^0 \to \omega f_2$ | $3.5 \times 10^{-11}$ | - | - |
| $\bar{B}_s^0 \to \omega f_2'$ | $2.6 \times 10^{-9}$ | - | - |
| $\bar{B}_s^0 \to \phi f_2$ | $5.6 \times 10^{-10}$ | - | - |
| $\bar{B}_s^0 \to \phi f_2'$ | $4.1 \times 10^{-8}$ | - | - |
| $\Delta b = 1, \Delta C = -1, \Delta S = 0$ | | | |
| $B^- \to D^{*-} a_2^0$ | $9.6 \times 10^{-7}$ | - | - |
| $B^- \to D^{*-} f_2$ | $1.0 \times 10^{-6}$ | - | - |
| $B^- \to D^{*-} f_2'$ | $2.1 \times 10^{-8}$ | - | - |
| $B^- \to \bar{D}^{*0} a_2^-$ | $1.1 \times 10^{-7}$ | - | - |
| $\bar{B}^0 \to D^{*-} a_2^+$ | $1.8 \times 10^{-6}$ | - | - |
| $\bar{B}^0 \to \bar{D}^{*0} a_2^0$ | $5.0 \times 10^{-8}$ | - | - |
| $\bar{B}^0 \to \bar{D}^{*0} f_2$ | $5.3 \times 10^{-8}$ | - | - |
| $\bar{B}^0 \to \bar{D}^{*0} f_2'$ | $1.1 \times 10^{-9}$ | - | - |
| $\bar{B}_s^0 \to D^{*-} K_2^+$ | $1.2 \times 10^{-6}$ | - | - |
| $\bar{B}_s^0 \to \bar{D}^{*0} K_2^0$ | $7.0 \times 10^{-8}$ | - | - |